\documentstyle[12pt,epsfig]{article}
 \newcommand{\beq}{\begin{equation}}
 \newcommand{\be}{\begin{eqnarray}}
 \newcommand{\eeq}{\end{equation}}
 \newcommand{\ee}{\end{eqnarray}}
 \newcommand{\ba}{\begin{array}{1}}
 \newcommand{\ea}{\end{array}}
 \newcommand{\bb}{\begin{thebibliography}}
 \newcommand{\eb}{\end{thebibliography}}
 \newcommand{\ci}[1]{\cite{#1}}
 \newcommand{\bi}[1]{\bibitem{#1}}

 \newcommand{\abstitle}[1]{{\small{\bf #1}}}

\begin{document}
 \begin{center}
 \abstitle{HADRON MULTIPLICITY IN LEPTON-NUCLEON
 INTERACTIONS} \\
 \vspace{0.5cm}
{G. I. Lykasov$^a$, U. Sukhatme$^b$,
 V. V. Uzhinsky$^a$}\\
 \end{center}
\vspace{0.5cm}
{$^a$ Joint Institute for Nuclear Research,
 Dubna, Moscow Region, 141980, Russia.} \\
{$^b$ State University of New York
 at Buffalo, 810 Clemens Hall, 14260-4600,
 USA.} \\
 \begin{center}
 {\bf Abstract}
 \end{center}

 Multi-hadron production in inelastic neutrino-nucleon
 interactions is investigated within the framework of the quark-gluon string model. The contributions of the planar
 (one-Reggeon exchange) and cylindrical (one-Pomeron exchange)
 graphs to different observables is computed using
 a Monte Carlo program for the generation of hadrons produced from the decay of
 colorless quark-antiquark  strings. The suggested approach results in a satisfactory description
 of the experimental data on
 $\nu(\bar\nu) N\rightarrow\mu^-(\mu^+) h X$
 reactions obtained recently at CERN by the NOMAD Collaboration. The data extends over a wide range of initial neutrino energies $E_\nu$ $<$ 200 GeV/c and momentum transfers 1 $<$ Q $<$ 7 GeV/c, well into the region where perturbative QCD calculations are not applicable.

 \vspace{0.4cm}

 \vspace{0.5cm}
 The investigation of multi-hadron production in inelastic lepton-nucleon
 interactions is a tool to study the dynamics of such processes
 and the nucleon structure function, particularly the $Q^2$ dependence over a wide interval of values.
 In analyzing these types of reactions, the fragmentation properties
 of the struck quark in deep inelastic scattering (DIS) in $\nu_\mu N$
 and ${\bar\nu}_\mu N$ charged current events are often compared to those of the
 quarks produced in $e^+e^-$ annihilation and $ep$ inelastic scattering.
 This comparison tests the universality of the quark fragmentation
 process in regions where non-perturbative QCD effects are important.
 Such an analysis has been performed recently in Ref. \ci{nomad1}. It has also been found that a perturbative QCD calculation results in a good description of the charged hadron multiplicity $<n_{ch}>$ measured in $e^+e^-$ annihilation and $ep$ DIS
 at initial energies E = $\sqrt{s}$ = Q $>$ 5 GeV/c \ci{nomad1}, whereas a similar calculation of $<n_{ch}>$ in the current region of
 processes $\nu(\bar\nu)~p\rightarrow\mu^-(\mu^+)~h~X$ cannot
 reproduce the NOMAD data at Q $<$ 5 GeV/c .
 This discrepancy is an important motivation for the present analysis of inelastic neutrino-nucleon
 interactions within the framework of the quark-gluon string model (QGSM) \ci{qgsm1,qgsm2}, which is essentially equivalent to the
 dual parton model (DPM) developed in Refs. \ci{dpm1,dpm2}.
 We calculate the different observables in a manner similar to Ref. \ci{bfls}, which focused on the
 pion multiplicity in  $\nu(\bar\nu)~A\rightarrow\mu^-(\mu^+)~h~X$
 reactions. The main difference from the previous study is
 that we now use the Monte Carlo (MC) generation of hadrons produced
 from the decay of colorless
 strings. This procedure allows us to analyze all the observables
 measured experimentally and make a comparison with recent data \cite{nomad1}.

 In the QGSM \ci{qgsm1,qgsm2}, hadron production in the reactions
 $\nu(\bar\nu)~p\rightarrow\mu^-(\mu^+)~h~X$ is
 described in terms of planar and cylindrical graphs, as shown in
 Fig. 1. The planar graph of Fig. 1(a) describes neutrino scattering off a
 valence quark, corresponding to the one Reggeon exchange in the
 $t$-channel \ci{qgsm1}, whereas the cylindrical graph of Fig. 1(b)
 describes neutrino scattering off sea quarks in the proton
 corresponding to the one-Pomeron exchange in the $t$-channel
 \ci{qgsm1}. The figures also show the occurrence of hadronization in the colorless quark-antiquark and quark-diquark strings.

 The relativistic invariant distribution of hadrons produced in
 the process $\nu(\bar\nu)~p\rightarrow\mu^-(\mu^+)~h~X$
 is defined as
 \be
 \rho_{\nu(\bar\nu)~p\rightarrow\mu^-(\mu^+)~h~X}~=~
 E_h\frac{dN}{d^3p_hd\Omega dE^\prime}~~,
 \label{def:dN}
 \ee
 where $E_h$ and ${\bf p}_h$ are the total energy and three
 momentum of the produced hadron, respectively; $E^\prime$ and $\Omega$
 are the energy and the solid angle of the final muon.
 It can be written in the following general form
 \be
 \rho_{\nu(\bar\nu)~p\rightarrow\mu^-(\mu^+)~h~X}~=~
 \Phi(Q^2)\left\{F_P(x,Q^2;z,p_{ht})~+~F_C(x,Q^2;z,p_{ht})
 \right\}
 \label{def:rho}
 \ee
 with
 $$
  \Phi(Q^2)~=~mE\frac{G^2}{\pi}\frac{m_W^2}{Q^2+m_W^2}~,
 $$
 where $G$ is the Fermi weak coupling constant, $E$ is the energy
 of the incoming neutrino, $m$ and $m_W$ are the nucleon and the
 $W$-boson masses respectively, $x=Q^2/2(p_\nu\cdot k)$ is
 the Bjorken variable, $p_\nu$ and $k$ are the four momenta of
 the initial neutrino and nucleon. Here
 $F_P(x,Q^2;z,p_{ht})$ and $F_C(x,Q^2;z,p_{ht})$ are the
 probability distributions of produced hadrons corresponding
 to the planar and cylindrical graphs of Fig. 1
 respectively. The contributions $F_P$ and $F_C$ are computed
 by using the MC generation of all the quark-antiquark and
 quark-diquark strings drawn in Fig. 1(a) and 1(b).
 Generally, $z$ is the light cone variable defined as
 $z=(E_h+p_{hz})/(E+p_z)$, where $p_{hz}$ is the hadron longitudinal momentum
 respect to the initial neutrino having momentum $p_z$, see fro example
 Refs.\ci{qgsm1,qgsm2} and \ci{dpm1,dpm2}. At large energies of final hadrons it is
 the longitudinal momentum fraction of hadron with respect to the neutrino in the
 rest frame of the proton-target. The variable $z$ can be treated also as the Feynman
 variable $x_F=2p_L^*/W_X$ defined as the longitudinal momentum fraction in the hadronic
 center of mass system (h.c.m.s.). Here $p_L^*$ is the longitudinal
 hadron  momentum in the h.c.m.s., $W_X$ is the mass of hadrons
 produced in the reaction, then $p_{ht}$ is the transverse momentum
 of produced hadron with respect to the current (hadronic jet)
 direction, see for example Ref. \ci{nomad2}.

 The main ingredients for the calculations of observables in the
 discussed reaction are the quark distributions in a nucleon and
 their fragmentation functions to hadrons. In addition to the dependence upon $x$, quark
 distributions also depend on $Q^2$ and the transverse momentum
 $k_t$. Following Ref. \ci{bfls}, we take a
 factorized form for these distributions:
 \be
 q_f(x,Q^2;k_t)~=~ q_f(x,Q^2)g_q(k_t)~,
 \label{def:xqkt}
 \ee
 with the function $g_q$ chosen in the form
 \be
 g_q(k_t)~=~\frac{B^2}{2\pi}e^{-Bk_t}
 \label{def:gkt}~,
 \ee
 where $B=1/<k_t>\simeq 4 (GeV/c)^{-1}$, and
 $<k_t>\simeq 0.25 GeV/c $ is the average transverse
 momentum of a quark in a nucleon.
 As for the function  $q_f(x,Q^2)$ we use
the fit suggested in Ref. \ci{ckmt} including the true
 Regge $x$-asymptotic at $x\rightarrow 0,
 x\rightarrow 1$ and small $Q^2$, and its QCD prediction
 at large $Q^2$. More specifically,  the valence $u$ and $d$ quark
 distributions according to Ref. \ci{ckmt}, have the
 following forms:
 \be
 q_u(x,Q^2)~=~B_ux^{-\alpha_R(0)}(1-x)^{n(Q^2)}
 \left(\frac{Q^2}{Q^2+b}\right)^{\alpha_R(0)}~,
 \label{def:qu}
 \ee
 \be
  q_d(x,Q^2)~=~B_dx^{-\alpha_R(0)}(1-x)^{n(Q^2)+1}
 \left(\frac{Q^2}{Q^2+b}\right)^{\alpha_R(0)}~,
  \label{def:qd}
 \ee
 where
 $$
 n(Q^2)~=~=\frac{3}{2}\left(1~+~\frac{Q^2}{Q^2+c}\right)~.
 $$
We take the values of the constants from Ref. \ci{ckmt}: $\alpha_R(0)=0.4150$ is the Reggeon intercept, $B_u=1.2064,
 B_d=0.1798, b=0.6452, c=3.5489$ .

 The sea quark distribution in the proton is taken to be \ci{ckmt}:
 \be
 q_{sea}(x,Q^2)~=~Ax^{-\Delta(Q^2)-1}(1-x)^{n(Q^2)+4}
 \left(\frac{Q^2}{Q^2+a}\right)^{1+\Delta(Q^2)}~,
 \label{def:qsea}
 \ee
 where
 $$
 \Delta(Q^2)~=~\Delta_0\left(1+\frac{\Delta_1(Q^2)\times Q^2}
 {Q^2+\Delta_2}\right)~,
 $$
 and
 $A=0.1502, a=0.2631, \Delta_0=0.07684, \Delta_1=2.0,
 \Delta_2=1.1170$.
 Note, this fit of the quark distributions in the proton
 describes all the experimental data from very small $x$ up to
 $x\sim 0.9$ \ci{ckmt}.

 Generally, the fragmentation functions (FF) of quarks
 (diquarks) $D_{q(qq)}^h$ into hadrons depend on the hadron
 momentum fraction $z_1$ and the hadron transverse momentum
 ${\tilde p}_{ht}$ with respect to a quark (diquark) momentum
 direction. Here also we choose the factorized form
 \be
 D_{q(qq)}^h(z_1,{\tilde p}_{ht})~=~ D_{q(qq)}^h(z_1)g_q({\tilde p}_{ht})~,
 \label{def:Dq}
 \ee
 with the function $g_q$ again chosen in the form
 (\ref{def:gkt}).
 The functions $D_{q(qq)}^h(z_1)$ are constructed, according to
 the recursive cascade model procedure suggested in \ci{ff1}. They are found from the
 following integral equation \ci{ff1}:
 \be
 D_{q(qq)}^h(z_1)~=~
 f(z_1)~+~\int_{z_1}^1f(x) D_{q(qq)}^h\left(\frac{z_1}{x}
 \right)
 \frac{dx}{x}
 \label{def;frf}
 \ee
 The function $f(x)$ is chosen in the form
 \be
 f(x)~=~x^\beta(1-x)^\gamma
 \label{def:fx}
 \ee
 According to the main points of the QGSM, the fragmentation functions
 $D_{q(qq)}^h(z_1)$ should satisfy true Regge asymptotics
 at $z_1\rightarrow 1$ and $z_1\rightarrow 0$ \ci{qgsm2}.
 This limitation allows us to find the values of the parameters.
 The detailed procedure is presented in Ref. \ci{shmuzh}.
 In general case the FF have to depend not only on $z_1,p_{ht}$
 but also on $Q^2$. At small $Q^2$ they have to reproduce the true
 Regge asymptotic \ci{qgsm2} and at large $Q^2$ they have to describe
 the $e^+e^-$ annihilation data. Since we analyze inelastic neutrino-
 nucleon interactions mainly at not large $Q^2$, one can assume that
 the FF depends on it weakly and neglect this $Q^2$ behaviour of
 $D_{q(qq)}^h$.

 The MC computation results of the different observables in the
 reaction $\nu+N\rightarrow\mu^-+h+X$ are presented in Figs. 2 to 5.
 The mean charged multiplicity in the current region
 is presented in Fig. 2. As it shown in \ci{nomad1} the NOMAD
 multiplicity values $n_{ch}$ are very close to $<n_{ch}>/2$
 results from $e^+e^-$ experiment \ci{jade} at energy $E=\sqrt{s}$
 and $<n_{ch}>$ from $ep$ and ${\bar\nu}_\mu p$ at $E=Q$.
 In Fig. 2 we compare our results with the QCD calculation of the
 charged multiplicity from $e^+e^-$ $<n_{ch}^{QCD}>/2$. The open
 circles in Fig. 2 correspond to the QCD calculations \ci{QCD1,QCD2}
 for the evolution of partons in the leading log approximation
 which give for $<n_{ch}^{QCD}>$ the following fit, see also
 \ci{nomad1,tasso},
 \be
 n_{ch}^{QCD}~=~a+b\,\exp(c\sqrt{ln(Q^2/Q_0^2)}) ,
 \label{def:nchqcd}
 \ee
 where
 $a=2.257, b=0.094, c=1.775, Q_0=1 GeV/c$. The solid, long
 dash and short dash lines in Fig. 2 correspond to our Monte Carlo
 calculation at the initial neutrino energies $E_\nu=150 GeV,
 E_\nu=45 GeV, E_\nu=23.6 GeV$ respectively. The NOMAD
 experimental data are averaged over the initial energy
 \ci{nomad1}. It is seen from Fig. 1 that the QCD fit
 can't reproduce the NOMAD data on the charged multiplicity
 at $Q < 5 GeV/c$, whereas at large $Q$ it describes these ones
 and $<n_{ch}>$ from $e^+e^-$, $ep$ and ${\bar\nu}_\mu p$
 experiments satisfactory as it is shown in \ci{nomad1}. The suggested
 approach describes the NOMAD data at $Q~<~5 GeV/c$ rather well.

 In Fig. 3 the $Q^2$ dependence of the inclusive muon spectrum
 in the process $\nu+p\rightarrow\mu^-+X$ is presented. The long
 and the short dash lines correspond to the the contributions
 of the cylindrical (one-Pomeron exchange) and the planar
 (one-Reggeon exchange) graphs to this spectrum, the solid line
 shows their incoherent sum. It is seen from Fig. 3 that the
 contribution of the one-Pomeron exchange graph is decreasing
 with increase of $Q^2$ much faster than the one of the
 one-Reggeon diagram which dominates at $Q^2>20 (GeV/c)^2$.

 In Fig. 4 the $x_F$ distributions of strange $K^0$-mesons and
 $\Lambda$-baryons are presented. The solid lines in Fig. 3
 correspond to our calculation, the experimental points are
 the NOMAD data \ci{nomad2}. This figure shows rather
 satisfactory description of the NOMAD data.

 In Fig. 5 our calculation results of the mean multiplicity of
 the strange hadrons $K_S^0,\Lambda$ and $\bar\Lambda$ as the
 function of the neutrino energy $E_\nu$ with the NOMAD data
 \ci{nomad2} are presented.

 In this Letter we have analyzed the multiple hadron production
 in inelastic neutrino-nucleon processes, The main conclusions
 of our work can be summarized as follows. By the analysis of the
 charged hadron multiplicity in the current region of the process
 $\nu+p\rightarrow\mu^-+h+X$ at $Q^2\leq 10 (GeV/c)^2$ the
 nonperturbative
 QCD effects become very important. The conventional perturbative QCD
 calculation of $<n_{ch}>$ doesn't reproduce the NOMAD data in this
 region, whereas the application of the QGSM or the DPM allows to
 describe these data rather satisfactory. The main contribution
 to this observable and the muon distribution  $1/Nd\sigma/dQ^2$ at
 small $Q^2$ is coming from the cylindrical (one-Pomeron exchange)
 graph, Fig. 1(b), whereas at large $Q^2$ the planar (one-Reggeon)
 graph, Fig. 1(a), dominates. Note, these two observables presented in
 Figs. 2,3 are related to the current fragmentation region. Recently in
 \ci{bfls} it has been shown that in the target fragmentation region
 these two contributions to the multiplicity of pions produced in the
 backward hemi-sphere in inelastic neutrino-nucleus collisions have
 the $Q^2$ dependences opposite to the ones obtained in this paper.
 There is some analogous between the last reaction and the soft
 $h-N$ processes at $x_F\rightarrow -1$ when the one-Pomeron exchange
 graph dominates at large initial energy, see for example \ci{qgsm1}.
 Therefore the contributions of the Fig. 1(a) and Fig. 1(b) graphs have
 different $Q^2$ dependences in the different kinematical regions, the
 current and the target fragmentation ones.

 The NOMAD data on the $x_F$ distributions of strange hadrons are
 described by the QGSM rather satisfactorily. It can mean
 that the fragmentation functions of quarks (diquarks) used by description
 of soft $h-N$ processes are rather good to analyse the multiple hadron production
 in inelastic $\nu-N$ interactions at not large $Q^2$.

 \vspace{1.cm}

 \noindent {\bf Acknowledgements} \\
 We are indebted to A. B. Kaidalov, B. Popov and A. Capella for many
 helpful discussions. One of the authors (V.V.U.) thanks RFBR grant
 N 01-02-16431 and INTAS grant N 00-00366 for their financial
 support. Support from the U.S. Department of Energy is gratefully acknowledged.

\newpage
\begin{figure}[t]
\begin{center}
\psfig{file=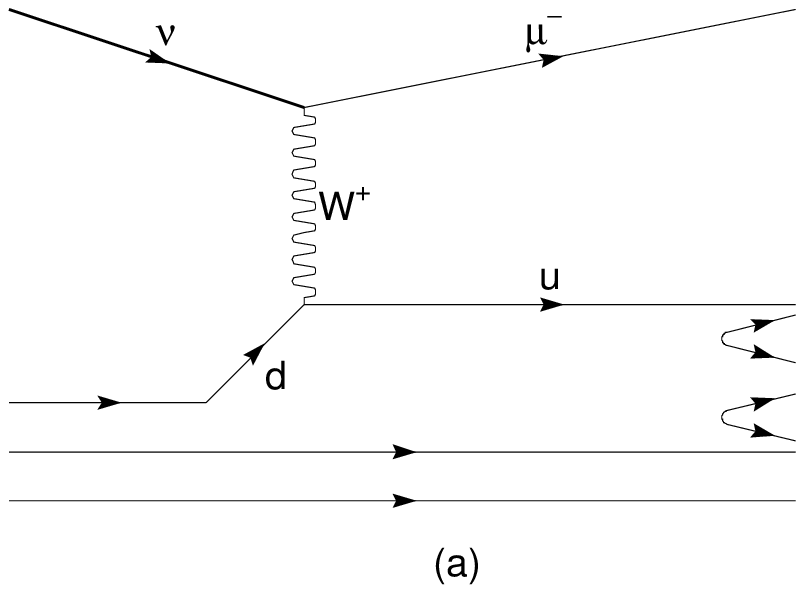,width=65mm,height=65mm,angle=0}\quad\quad\quad
\psfig{file=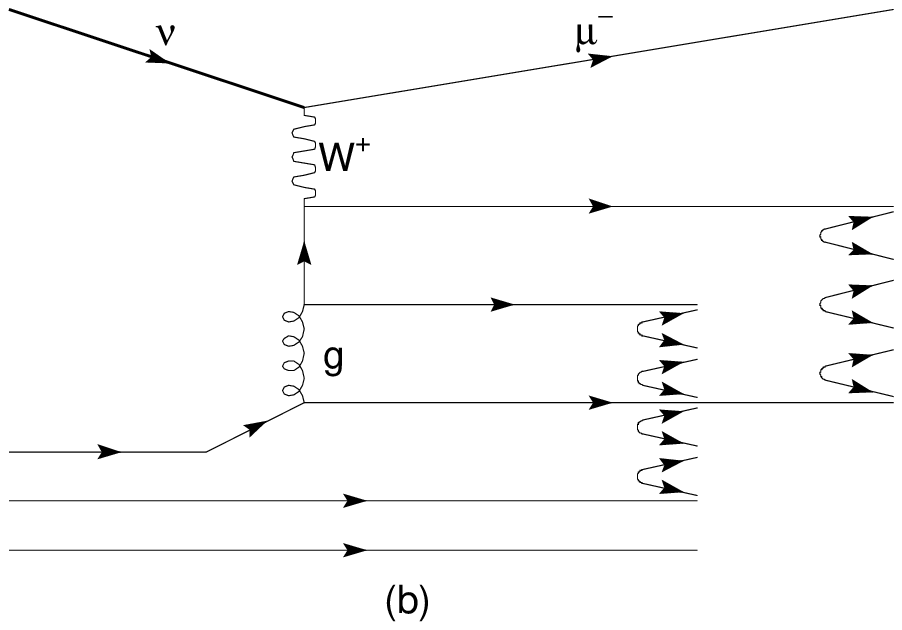,width=65mm,height=65mm,angle=0}
\end{center}
\caption[fig1a]{(a) The planar (one-Reggeon exchange) graph and (b)
the cylindrical (one-Pomeron exchange) graph.}
\end{figure}

 \begin{figure}[b]
 \psfull
 \begin{center}
 \epsfig{file=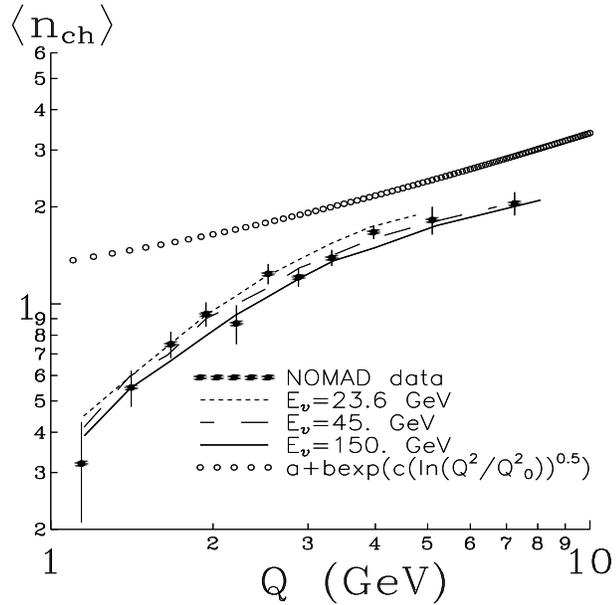,height=8.cm,width=8.cm}
 \end{center}
 \caption[fig2]{Mean multiplicity of charged hadrons
 in the current fragmentation region as a function of the momentum transfer $Q$.
 The open circles correspond to the QCD fit given by eq.(\ref{def:nchqcd});
 the solid, long dash and short dash lines correspond to our calculations at
 $E_\nu=150.Gev, 45.Gev.$ and $E_\nu=23.6GeV$ respectively.
 The experimental points are the NOMAD data \ci{nomad1}}
 \label{fig2}
 \end{figure}

 \begin{figure}[t]
 \psfull
 \begin{center}
 \epsfig{file=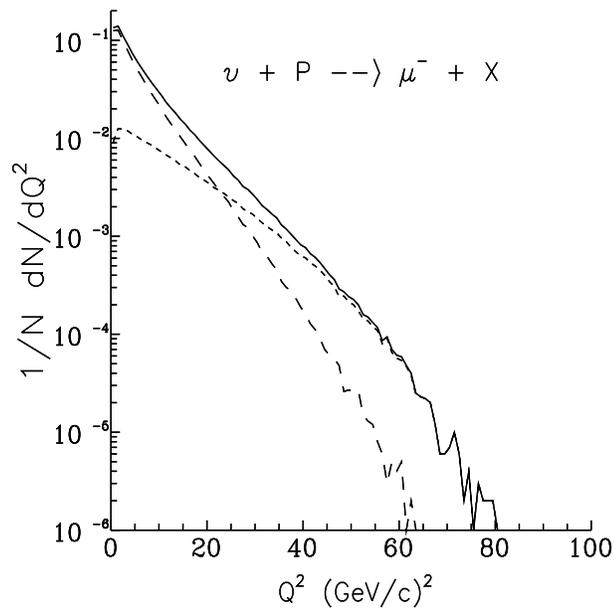,height=8.cm,width=8.cm}
 \end{center}
 \caption[fig3]{The $Q^2$-distribution $1/NdN/dQ^2$ of
 muons produced in the inclusive process
 $\nu+p\rightarrow\mu^-+X$. The long dash and short dash lines correspond
 to the contributions of the cylindrical and planar graphs respectively. The
 solid line is the sum of these contributions.}
 \label{fig3}
 \end{figure}

 \begin{figure}[b]
 \psfull
 \begin{center}
 \epsfig{file=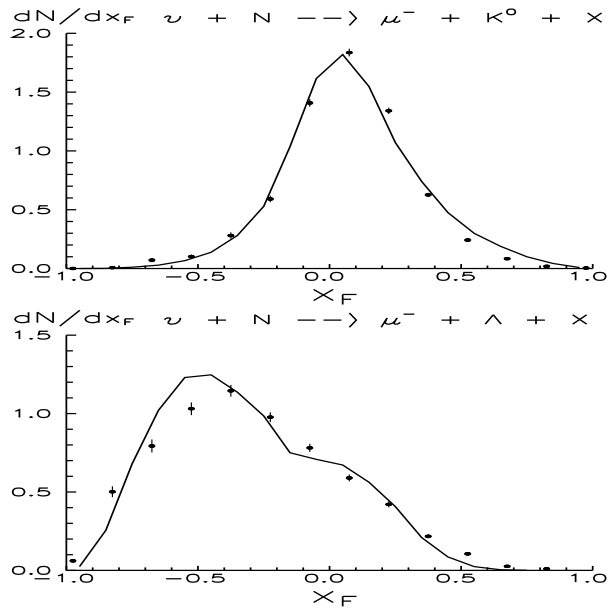,height=8.cm,width=8.cm}
 \end{center}
 \caption[fig4]{The $x_F$ distribution of strange hadrosn.
 The upper panel corresponds to the distribution $dN/x_F$  of
 $K^0$ mesons in semi-inclusive process  $\nu+N\rightarrow\mu^-+K^0+X$,
 whereas the down panel shows the $x_F$ distribution of $\Lambda^0$ hyperons
 produced in $\nu+N\rightarrow\mu^-+\Lambda^0+X$ reaction. The experimental points
 are the NOMAD data \ci{nomad1}. }
 \label{fig4}
 \end{figure}

 \begin{figure}[ht]
 \begin{center}
 \psfig{file=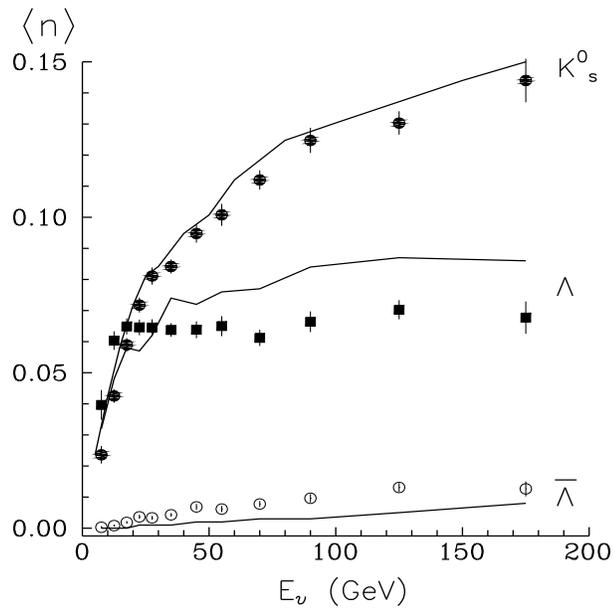,height=8.cm,width=8.cm,angle=0}
 \end{center}
 \caption[fig5]{The average multiplicity of the strange
 hadrons as a function of the neutrino energy $E_\nu$. The upper line corresponds
 to $<n>$ of the $K_s^0$ mesons, the middle line is the mean multiplicity of
 $\Lambda^0$ hyperons and the down curve shows $<n>$ of the antilambda hyperons
 ${\bar\Lambda}^0$.}
 \label{fig5}
 \end{figure}

 \end{document}